\documentstyle[12pt]{article}
\title{Position Measurement\\ for a Relativistic Particle:\\
   Restricted-Path-Integral Analysis\thanks{Published in Phys. Lett. 
A~208, 269-275 (1995).}}
\author{
Michael B. Mensky$^{\dag\ddag}$
\and
Horst v. Borzeszkowski$^{\ddag}$
\\[10pt]$^{\dag}$
P.N.Lebedev Physics Institute, 117924 Moscow, Russia\\
Email: mensky@mbm.fian.msk.su
\\[3pt]$^{\ddag}$
Institut f\"ur Theoretische Physik,
Technische Universit\"at Berlin\\
Hardenbergstr. 36,
D-10623 Berlin, Germany}
\date{}

\newcommand{\be}{\begin{equation}}
\newcommand{\ee}{\end{equation}}
\newcommand{\ba}{\begin{eqnarray}}
\newcommand{\ea}{\end{eqnarray}}
\newcommand{\ban}{\begin{eqnarray*}}
\newcommand{\ean}{\end{eqnarray*}}
\newcommand{\ra}{\rangle}

\newcommand{\al}{\alpha}

\newcommand{\tr}{{\rm Tr}}

\renewcommand{\a}{{\bf a}}
\renewcommand{\b}{{\bf b}}

\newcommand{\p}{{\bf p}}
\renewcommand{\l}{{\bf l}}
\newcommand{\Da}{\Delta a}
\newcommand{\x}{{\bf x}}
\renewcommand{\S}{{\cal S}}
\newcommand{\dzero}{\stackrel{\leftrightarrow}{\partial_0}}

\begin{document}
\maketitle
\abstract{Measurements of the position of a relativistic particle
is considered in the framework of the Restricted-Path-Integral
(RPI) approach. The amplitude describing such a measurement is
shown to be exponentially small outside the light cone of the
space-time point corresponding to the measurement output,
in a qualitative agreement with the Hellwig and Kraus' postulate
of relativistic state reduction. Theory of the measurement including
the probability distribution for different measurement outputs
is suggested. It is shown that correct theory does not exist (for
arbitrary initial states) if the error $\Da$ of the measurement is
less than the Compton length $\lambda_C=\hbar/mc$. The physical reason
is that the picture of measurement is destroyed in this case by pair
creation.}

\section{Introduction}\label{intro}

In nonrelativistic quantum mechanics, measurements are described
by the von Neumann's postulate of the state reduction (wave
function collapse). In relativistic theory this is impossible
since an instantaneous state reduction contradicts to causality
and therefore must be modified. This problem has been considered
by many authors (see for example
\cite{Hellw}-\cite{Finkelst94},
but consensus was not achieved. We
shall consider the measurement of the position of a relativistic
particle with the help of the Restricted-Path-Integral (RPI)
approach to continuous quantum measurements
\cite{M79,book93}. The results will be shown to be in a
qualitative agreement with the Hellwig and Kraus' postulate
\cite{Hellw} according to which the state reduction occurs in the
light cone of the of the measurement event.

Restricted-Path-Integral (RPI) approach has been proposed by
R.Feynman \cite{Feynman48} for description of continuous
(prolonged in time) measurements and technically elaborated in
\cite{M79,book93} (see also \cite{RPI-other}). The idea of the
approach is that evolution of the system undergoing a continuous
measurement must be described by the path integral restricted on the
set of paths compatible with the measurement output.

The approach proved to be effective for different types of
measurements of non-relativistic systems as well as for
measurements of relativistic quantum fields (electromagnetic and
gravitational, see \cite{book93,emHorst}). Its advantage is in
model-independence and generality. In the present paper the RPI
approach will be applied to the problem of measurement of position of
a relativistic particle.

Let the position of a relativistic particle be measured at a
specified time moment and the measurement output correspond to
the point (event) $a$ belonging to the corresponding time slice.
Then the paths $[x]$ compatible with the output $a$ are those
crossing this point, $a\in [x]$. Therefore, only these paths
contribute the evolution of a particle subject to the
measurement.

The problem is therefore reduced to calculating relativistic path
integrals over the sets of paths crossing the given space-time
point. Technical difficulties of this calculation will be
overcome due to the specific properties of the causal propagator
(the path integral over all paths).

\section{Relativistic Path Integrals}\label{path-relat}

The causal propagator (transition amplitude) for a relativistic
particle can be expressed in the form of a path integral if one
introduces, following Stueckelberg \cite{Stueck}, the fifth
parameter (besides four space-time coordinates) $\tau$ called
{\em proper time} or {\em historical time}.

Consider for simplicity a scalar particle of the mass $m$. Its
{\em causal propagator} is equal to the integral over the
proper time,\footnote{we shall use in the present paper the
natural units $\hbar=c=1$ everywhere but in
the discussion of the results}
\be
K(x''-x')=\int_0^{\infty}d\tau\,
   \exp\left( -i(m^2-i\epsilon)\tau\right)
   \,K_{\tau}(x''-x'),
\label{tau-int}\ee
of a subsidiary {\em proper-time-dependent propagator}. The
latter, in turn, may be given the form of a path integral:
\be
K_{\tau}(x''-x')=\int_{x''\leftarrow x'} d[x]_{\tau}
  \, \exp\left( -\frac{i}{4}
  \int_0^{\tau}(\dot x,\dot x)d\tau\right).
\label{path-int}\ee
Here $(,)$ denote the Lorentzian inner product and the usual
definition of the measure is taken (see for example \cite{book93} for
elementary definitions from theory of path integrals).

As a result of these definitions, the subsidiary
proper-time-dependent propagator satisfies the relativistic
Schr\"odinger-type equation
\be
\frac{d}{d\tau}K_{\tau}(x''-x')
  = -i\, \Box K_{\tau}(x''-x')
\label{rel-Schroed}\ee
and the causal propagator $K_{\tau}(x''-x')$ is a Green function of
the Klein-Gordon equation.

For the analysis of continuous measurements of a relativistic
particle in the framework of the Restricted-Path-Integral (RPI)
approach we have to deal with the path integrals of the type of
Eqs.~(\ref{tau-int}),~(\ref{path-int}) but restricted on the sets
of paths compatible with the corresponding measurement outputs.

\section{The Measurement Amplitude}\label{amplit-posit}

We shall consider measurement of the particle position at time
moment $x^0=ct$. First the overidealized situation of an
absolutely precise measurement will be analyzed. A finite
measurement error will be taken into account later on
(Sect.~\ref{fin-error}).

If the position of the particle is precisely measured at time
$t$, then the measurement outputs may be described as
three-vectors $\a$ or as points $a=(ct,\a)$  of the time slice
$t={\rm const}$ of the space-time, i.e. points of the space-like
surface $\S=\{ x | x^0=ct\}$.

If we know that the measurement of the position (at time $t$) has
given the result $\a$, then we know that the world line
(trajectory) of the particle crossed the surface $\S$ in the
point $a$. Therefore, instead of the integral over all paths,
evolution of the particle must be described by the integral over
the set $I_a$ of paths crossing $\S$ in the point $a$.

The set of paths $I_a$ is a ``corridor'' describing adequately
the measurement output. In the Restricted-Path-Integral method
restricting of the path integral
(\ref{tau-int}),~(\ref{path-int}) onto this corridor will give an
amplitude of transition under the measurement, or the {\em
measurement amplitude}:
\ba
K^{(a)}(x'',x')&=&\int_0^{\infty}d\tau\,
   \exp\left( -\frac{i}{\hbar}
     (m^2-i\epsilon)\tau\right)
   \,K_{\tau}^{(a)}(x'',x'), \nonumber\\
K_{\tau}^{(a)}(x'',x')&=&\int_{x''\leftarrow a\leftarrow x'} d[x]_{\tau}
  \, \exp\left( -\frac{i}{4}
  \int_0^{\tau}(\dot x,\dot x)d\tau\right).
\label{pos-RPI}\ea

This restricted path integral is evidently connected in some way
with the product of two unrestricted integrals of the type of
Eqs.~(\ref{tau-int}),~(\ref{path-int}), one integral from the
point $x'$ to the point $a$, the other from $a$ to $x''$:
\be
K^{(a)}(x'',x')= K(x''-a) * K(a-x').
\label{pos-prod}\ee
The precise definition of this product is to be found.

To find the correct definition for the product (\ref{pos-prod}),
we shall require that the set of the amplitudes (\ref{pos-prod})
be complete. This means that summation of the amplitudes
corresponding to all possible values of $\a$ should give the
amplitude describing the evolution without measurement:
\be
\int_{\S} d^3\a \, K^{(a)}(x'',x')= K(x''-x').
\label{unity-decomp}\ee

This relation, with the expression (\ref{pos-prod}) in the integrand,
resembles the known Kolmogorov-type property of the propagator,
\be
i\int_{x^0=ct} d^3\x \,K(x''-x) \dzero K(x-x')= K(x''-x')
\label{Kolmogor-flat}\ee
where $x''^0>ct>x'^0$ and it is denoted
$$
f(x)\dzero g(x)=
f(x)\stackrel{\leftrightarrow}{\frac{\partial}{\partial x^0}}g(x)=
f(x)\frac{\partial g(x)}{\partial x^0}
-\frac{\partial f(x)}{\partial x^0}g(x).
$$
Therefore, the completeness of the measurement amplitudes will be
provided if we define the product (\ref{pos-prod}) as
follows:\footnote{This amplitude will be used only for estimating
relative probabilities, so that its normalization is not
important. We shall consider the normalization in
Sect.~\ref{fin-error}, discussing the measurement with a finite
error.}
\be
K^{(a)}(x'',x')=
i\, K(x''-a)
\stackrel{\leftrightarrow}{\frac{\partial}{\partial a^0}}
K(a-x').
\label{precise-meas-ampl}\ee

We found the amplitude (\ref{precise-meas-ampl}) requiring that
summation of such amplitudes with different $a$ gives the
propagator of a free particle (without any measurement). This
requirement is natural in the framework of the theory of free
particles because of the superposition principle. After this,
when the form of the amplitude (\ref{precise-meas-ampl}) is
found, we go over from theory of a free particle to theory of a
measured particle, where the role of these amplitudes will be
quite different because the superposition principle does not take
place.

In quantum theory of measurements the superposition principle is
restricted: the amplitudes corresponding to different measurement
outputs may not be added. In our case, when the precise measurement of
the position is under consideration, the amplitudes
(\ref{precise-meas-ampl}) with different $a$ correspond to different
measurement outputs. Therefore, each of them must be used separately,
and their summation has no sense. Amplitudes $K^{(a)}(x'',x')$ with
different $a=(ct,\a)$ are incoherent.\footnote{Later on we shall
consider the measurement with a finite precision. Then the amplitudes
with close $\a$ (differing less than by the measurement error) are
coherent and may be summed up (see Sect.~\ref{fin-error}).}

The amplitudes (\ref{precise-meas-ampl}) are derived for a particle
which is in the space-time point $x'$ before the measurement and in
the point $x''$ after it. The realistic situation corresponds usually
to the initial and final states given by the wave functions at the
corresponding time moments $t'$, $t''$. The measurement amplitudes are
then
\be
K^{(a)}(\psi'',\psi')
  =-\int d^3\x' \,d^3\x'' \,
  \overline{\psi''^(x'')}\,
\stackrel{\leftrightarrow}{\frac{\partial}{\partial x''^0}}
  \,K^{(a)}(x'',x')\,
\stackrel{\leftrightarrow}{\frac{\partial}{\partial x'^0}}
  \,\psi'(x').
\label{pos-wavef}\ee
where the bar denotes a complex conjugate.

\section{Properties of the Amplitude}\label{precise-meas}

So far we talked about the amplitude (\ref{precise-meas-ampl}) in such
a way as if it described measurement of the position with absolute 
precision. In other words, the measurement described by this
amplitude was supposed to consist in localizing the particle in a 
single point $a=(ct,\a)$ on the surface $\S$. We shall see later 
(Sect.~\ref{sect-gen-unitar}) that the only measurements which may be 
described correctly are those with a finite (and not too small) error. 
The amplitude (\ref{precise-meas-ampl}) cannot be correctly 
interpreted. Nevertheless, it is important to investigate the 
properties of this amplitude. The properties of the finite-error 
amplitudes will follow then straightforwardly leading to physical 
conclusions.

For simplicity, we shall use physical terms in the analysis of the 
amplitude (\ref{precise-meas-ampl}) as if it could be interpreted 
physically. In fact, the present section is devoted to investigation 
of {\em mathematical properties} of the subsidiary amplitude while 
{\em physical interpretation} is possible only for finite-error 
amplitudes of Sect.~\ref{fin-error}.

Consider therefore the amplitudes (\ref{precise-meas-ampl}) and 
(\ref{pos-wavef}) as those describing evolution of the particle 
undergoing the position measurement. Then they are transition amplitudes from 
the point $x'$ to the point $x''$ (correspondingly from the state 
$\psi'$ to the state $\psi''$) under the condition that the 
measurement carried out at time $t$ gave the output $\a$. 
The amplitudes allow one to evaluate the probability that the particle 
achieves a certain state given an initial state and a measurement 
output.

Instead of this, one may interpret the same amplitudes
(\ref{precise-meas-ampl}), (\ref{pos-wavef}) as the probability
amplitudes for different measurement outputs $\a$, given the
initial and final states ($x'$ and $x''$ or $\psi'$ and
$\psi''$). Relative probabilities of different measurement
outputs may be estimated as square modula of the amplitudes.
A mathematically rigorous definition of probabilities must 
include the generalized unitarity condition (see 
Sect.~\ref{sect-gen-unitar}). Some conclusions however may be made 
without this.

The first conclusion may be made directly from the form of
the amplitude (\ref{precise-meas-ampl}). The causal propagator
$K(x'',x')$ is exponentially small if the interval $x''-x'$ is
outside the light cone. Therefore, the amplitude
(\ref{precise-meas-ampl}) is small if the point $x'$ is outside the
past light cone of the point $a$ or/and $x''$ is outside the
future light cone of $a$.

This property is in the qualitative agreement with the postulate
of Hellwig and Kraus \cite{Hellw} that reduction of the wave
function of a relativistic system (for example a field) occurs
not at the moment of the measurement but in the light cone of the
space-time region where the measurement takes place (see also
\cite{AharonAlb81} for a critical discussion of this postulate).
Now we can derive the corresponding feature of the measurement
rather than postulate it.

The conclusion following from the mentioned property of the
amplitude (\ref{precise-meas-ampl}) may be formulated as follows:
\begin{itemize}
  \item The probability for the measurement to give the output
        $a$ is exponentially small if $a$ is not in the future
        light cone of the support of the initial wave function
        $\psi'$.
  \item The probability that the particle will be found in the
        state $\psi''$ after the measurement resulting in the
        output $a$, is exponentially small if $a$ is not in the past
        light cone of the support of $\psi''$.
\end{itemize}

In the above statements, `exponentially small' means decreasing
in $e$ times at distance of the order of the Compton length from
the boundary of the light cone.

The fact that the propagator does not abruptly disappear but
rather exponentially decreases outside the light cone seems to
contradict to causality, because the particle may seemingly be
discovered in the point that it cannot achieve by causal
evolution. However there is actually no contradiction. If one try
to demonstrate experimentally (even by a thought experiment) this
violation, one may see that such a demonstration is impossible
because of the uncertainty relation.

One more thing guaranteed by the same property of the propagator
(its exponential decreasing but not abrupt disappearing outside
the light cone) is that dependence of the amplitude
(\ref{precise-meas-ampl}) or (\ref{pos-wavef}) from the position
$\a$ is smoothed on scales of the order of the Compton length.
Therefore, probabilities of two measurement outputs $\a_1$ and
$\a_2$ are close if these outputs differ by the value of the
order of Compton length or less. Though we discuss the precise
measurement of position, the information (about the initial
state) given by this measurement cannot be more precise than up
to the Compton length. In the limits of the Compton length, the
measurement output may be arbitrary.

In a sense, the preceding argument means that the precise
measurement is impossible, the measurement error cannot be less
than the Compton length. In fact, we shall show below
(Sect.~\ref{sect-gen-unitar}) that the correct theory of the
measurement including a probability distribution exists for
arbitrary initial and final states only if the measurement is
performed with the error larger than the Compton length.

\section{Measurement with a Finite Error}\label{fin-error}

Consider now a measurement with a finite precision. Let the
measurement error be $\Da$. Then the measurement output $\a$
gives the information that the actual position of the particle
$\x$ differs from $\a$ not more than by the value $\Delta a$:
\be
|\x - \a | < \Da.
\label{Da-region}\ee
Such a measurement must be described by the amplitude
$$
K^{(a,\Da)}(x'',x')
=\int_{|\b - \a | < \Da} d^3\b\,K^{(b)}(x'',x').
$$

This interpretation of the measurement information corresponds to
a specific property of the measuring device. The information is
of this type if the device has a rectangular characteristic,
equal to unity in the region (\ref{Da-region}) and zero
otherwise.

In real situations measuring devices have smooth characteristics,
and the information supplied by the measurement is less definite.
If the measurement gives the output $\a$, this
means that an actual position of the particle with high
probability is very close to $\a$, with less probability is
somewhat further, and it is quite improbable that it differs from
$\a$ much more than by $\Da$. The amplitude describing such a
measurement has the form
\be
K^{(a,\Da)}(x'',x')
=\int d^3\b\, \rho(|\b - \a |) \, K^{(b)}(x'',x')
\label{smooth-ampl}\ee
with the weight function $\rho\ge 0$ concentrated in the region,
of the dimension $\Da$, around zero.

Again, just as in the case of the precise measurement, we should
use the amplitudes (\ref{smooth-ampl}) as incoherent ones. Each
of them describes evolution for a certain output of measurement.
Relative probabilities of different outputs may be roughly
estimated by the square modula of the corresponding amplitudes.
Mathematically rigorous concept of probabilities will be
discussed in Sect.~\ref{sect-gen-unitar}.

The analysis of Sect.~\ref{precise-meas} may be repeated with an
evident change for the finite-error measurements. Now we should
speak of the light cone of the region (\ref{Da-region}) around
the point $a$ rather than the light cone of a single point $a$.

\section{Generalized Unitarity}\label{sect-gen-unitar}

In the RPI approach to quantum continuous measurements
\cite{book93} evolution of the system undergoing the measurement
is described by a set of propagators $U_{\al}$ depending on
measurement outputs $\al$:
\be
|\psi_{\al}\ra = U_{\al} |\psi\ra, \quad
\rho_{\al} = U_{\al} \rho \left( U_{\al}\right)^{\dagger}.
\label{select-evolut}\ee
This is the evolution law for the {\em selective measurement}
when the measurement output is known. If it is unknown
({\em non-selective measurement}), then the density matrix after
the evolution is a sum of the density matrices corresponding to
all possible outputs:
\be
\rho'=\sum_{\al}\rho_{\al} =
\sum_{\al}U_{\al} \rho  {U_{\al}}^{\dagger}.
\label{non-select-evolut}\ee

Probability for the measurement output to belong to the set $A$
is equal to
$$
{\rm Prob}\,(\al\in A)=\sum_{\al\in A}\tr\, \rho_{\al}.
$$
Conservation of probabilities (normalization of $\rho'$) is
provided by the {\em generalized unitarity}
$$
\sum_{\al}
{U_{\al}}^{\dagger}U_{\al}={\bf 1}.
$$

In the case of {\em continuous set} of the measurement outputs
(typical for a continuous measurement) it is more correct to
speak about integration rather than summation over different
outputs. Particularly, the last formulas should be rewritten as
follows:
\ba
\rho'=\int d\mu (\al)\,\rho_{\al}, \quad
{\rm Prob}\,(\al\in A)
   =\int_{A}d\mu (\al)\,\tr\, \rho_{\al}
\label{evolut-measure}\\
\int d \mu (\al)\,
{U_{\al}}^{\dagger}U_{\al}={\bf 1}.
\label{unitarity-measure}\ea
The measure on the set of all outputs has to be chosen in such a
way as to provide the validity of the generalized unitarity
(\ref{unitarity-measure}).

We should now introduce the corresponding concepts
(probability and generalized unitarity) in our case.

The causal propagator (\ref{tau-int}) describes the evolution of
a (positive-frequency) state of a free particle:
\be
\psi(x'')=i\int_{\S'} d^3\x'\, K(x'',x')\dzero' \psi(x')
\label{free-evolut}\ee
where $\S'=\{ x'|x'^0={\rm const}\}$.
In the course of this evolution the inner product
\be
(\psi_1,\psi_2)
   = i \int_{\S} d^3\x\, \overline{\psi_1(x)}\dzero \psi_2(x)
\label{inner-prod}\ee
is conserved (here $\S=\{ x|x^0={\rm const}\}$). The evolution
described by the propagator $K(x'',x')$
is therefore unitary in the following sense:
\be
 i \int_{\S} d^3\x\, \overline{K(x,x'')}\dzero K(x,x')
  =K(x'',x').
\label{unitar}\ee

Eq.~(\ref{unitar}) represents unitarity of the theory
(conservation of probabilities) in an unusual way since there
are positive- and negative-frequency wave functions in
relativistic theory but we are interested only in
positive-frequency states of the particle. The meaning of
Eq.~(\ref{unitar}) is following. Acting by the propagator
$\overline{K(x,x'')}$ which is conjugate for $K(x,x'')$ we
describe propagation in an opposite direction, from the first
argument of the propagator to the second one. Therefore, the
action of $K(x,x')$ followed by the action of
$\overline{K(x,x'')}$ gives the same result as the action of
$K(x'',x')$.

If the arguments $x'$ and $x''$ belong to the same time slice, then
the action of the propagator $\overline{K(x,x'')}$ (according to
the formula (\ref{free-evolut})) does not change the wave
function. This brings us close to the usual form of unitarity. We
may suppose for simplicity that $x^0 > x''^0 > x'^0$. Then
positive-frequency part of the causal propagator may be used
instead of the complete propagator. The last two times $x''^0$
and $x'^0$ may be arbitrarily close to each other.

Unitarity (\ref{unitar}) of the causal propagator may be shown
to lead to the generalized unitarity
of the measurement amplitudes,
\be
i\int_{\S}d^3\a \,  \int_{\tilde\S} d^3\x \,
\overline{K^{(a,\Da)}(x,x'')}\dzero K^{(a,\Da)}(x,x')
  =K(x'',x'),
\label{gen-unitar-posit}\ee
if the error of the measurement (dimension of the region
where the function $\rho(|\b -\a|)$ is close to the maximum) is
larger than the Compton length $\lambda_C=\hbar/mc$.

More precisely, the condition for the generalized unitarity may be
formulated in terms of the Fourier expansion of the following
function:
\be
\int d^3\a \, \rho(|\b -\a|)\rho(|\b' -\a|)
 = \int d^3\l \; Q(\l)\; e^{i\l(\b' -\b)}.
\label{error-func}\ee
The generalized unitarity takes place if
\be
|\l| \ll \lambda_C =\frac{mc}{\hbar}
\label{Compton-condit}\ee
for all $\l$ in the region where the function $Q(\l)$ is not
negligible.

Besides this, the following normalization condition should be
fulfilled:
\be
\int d^3\l \, Q(\l) = 1.
\label{normalize}\ee

The equality (\ref{gen-unitar-posit}) is a concrete form of the
general relation (\ref{unitarity-measure}) in the coordinate
representation and with $d^3\a$ standing instead of $d\mu (\al)$.

The probability for the measurement output $a$ to belong to the
set $A\subset\S$ is expressed by the integral
\ba
{\rm Prob}\,(a\in A)
   = i \int_{A}d^3\a \,
    \int_{\S''}d^3\x'' \int_{\S'}d^3\x'_1 \int_{\S'}d^3\x'_2 \,
 \nonumber\\
   \times
    \overline{K^{(a,\Da)}(x'',x'_2)}
\stackrel{\leftrightarrow}{\frac{\partial}{\partial x''^0}}
    K^{(a,\Da)}(x'',x'_1)
\stackrel{\leftrightarrow}{\frac{\partial}{\partial {x'_1}^0}}
\stackrel{\leftrightarrow}{\frac{\partial}{\partial {x'_2}^0}}
    \rho (x'_1,x'_2).
\label{prob-posit}\ea
The generalized unitarity (\ref{gen-unitar-posit}) expresses
conservation of probabilities in the course of the evolution of
the system undergoing the position measurement.

The generalized unitarity does not take place if the measurement
of position is more precise than up to the Compton length. The
reason of this is evident. Localization of the particle in the
region of the dimension less than the Compton length requires
larger energy than the threshold of the pair creation. In this
case interaction of the particle with the localizing device has a
quite different character and cannot be described as a
measurement of position.

However, if we expand the relation (\ref{gen-unitar-posit}) in
the Fourier integral, we shall see that
only low-momentum components of this relation (with $|\p| <
\hbar/\Da$) are violated. Therefore, the description of the
measurement is correct for high-momentum states (with $|\p| \gg
\hbar/\Da$).

\section{Conclusion}\label{conclus}

We showed that the measurement of the position of a relativistic
particle can be correctly described in the framework of the
method of restricted path integrals. The results obtained are in
accordance with a more formal (not dynamical) consideration of the
measurements of this type (see \cite{Hellw} for the first attempt and
\cite{Finkelst94} and references therein for the recent papers).

It is interesting to apply the same method to other types of
measurements in relativistic systems, for example to measurement
of fields or to non-local measurements of different types. This
will be done in a separate paper.

\vspace{0.5cm}
\centerline{\bf ACKNOWLEDGEMENT}

The authors are indebted to
K.Hellwig for fruitful discussions of the
problem. This work was supported in part by the Deutsche
Forschungsgemeinschaft and the Russian Foundation for Fundamental
Research, grant 95-01-00111a.

\end{document}